\pdfoutput=1
\documentclass[twocolumn,aps,prl,preprintnumbers,amsmath,amssymb,superscriptaddress,nofootinbib]{revtex4-1}

\usepackage{amssymb,amsmath,bbold}
\usepackage{cancel,graphicx}
\usepackage[usenames,dvipsnames]{xcolor}
\usepackage[unicode=true,pdfusetitle,
 bookmarks=true,bookmarksnumbered=false,bookmarksopen=false,citecolor=Turquoise,
 breaklinks=false,pdfborder={0 0 1},backref=false,colorlinks=true,pdfpagemode=FullScreen]
 {hyperref}

\begin{document}

\preprint{IPMU17-0068}
\preprint{CERN-TH-2017-094}
\title{Anomaly-free local horizontal symmetry and anomaly-full rare $B$-decays}

\author{Rodrigo Alonso}
\email[]{rodrigo.alonso@cern.ch}
\affiliation{CERN, Theoretical Physics Department, CH-1211 Geneva 23, Switzerland}
\author{Peter Cox}
\email[]{peter.cox@ipmu.jp}
\affiliation{Kavli IPMU (WPI), UTIAS, University of Tokyo, Kashiwa, Chiba 277-8583, Japan}
\author{Chengcheng Han}
\email[]{chengcheng.han@ipmu.jp }
\affiliation{Kavli IPMU (WPI), UTIAS, University of Tokyo, Kashiwa, Chiba 277-8583, Japan}
\author{Tsutomu T. Yanagida}
\email[]{tsutomu.tyanagida@ipmu.jp}
\affiliation{Kavli IPMU (WPI), UTIAS, University of Tokyo, Kashiwa, Chiba 277-8583, Japan}
\affiliation{Hamamatsu Professor}

\begin{abstract}
The largest global symmetry that can be made local in the Standard Model + 3$\nu_R$ while being compatible with Pati-Salam unification is $SU(3)_H\times U(1)_{B-L}$.
The gauge bosons of this theory would induce flavour effects involving both quarks and leptons, and are a potential candidate to explain the recent reports of lepton universality violation in rare $B$ meson decays. 
In this letter we characterise this type of models and show how they can accommodate the data and naturally be within reach of direct searches.
\end{abstract}


\maketitle

\section{Introduction}

Lepton flavour universality (LFU) violation in rare $B$ meson decays provides a tantalising hint for new physics whose significance has recently increased~\cite{1705.05802}. A consistent picture may be beginning to emerge, with LHCb measurements~\cite{1406.6482, 1705.05802} of the theoretically clean ratios~\cite{hep-ph/0310219}
\begin{equation} \label{R_K}
  \mathcal{R}_K^{(*)}=\frac{\Gamma\left(B\rightarrow K^{(*)}\mu^+\mu^-\right)}{\Gamma\left(B\rightarrow K^{(*)}e^+e^-\right)} \,,
\end{equation}
in a combined tension of order $4\sigma$~\cite{1704.05435,1704.05438,1704.05340,1704.05444,1704.05447,Geng:2017svp} with the Standard Model (SM). 
Several phenomenologically motivated models have been proposed to explain this discrepancy (see \cite{1704.05438} for a review), one such possibility being a new $U(1)$ gauge symmetry~\cite{1310.1082, *1311.6729, *1403.1269, *1501.00993, *1503.03477, *1503.03865, *1505.03079, *1506.01705, *1507.06660, *1509.01249, *1510.07658, *1511.07447, *1601.07328, *1604.03088, *1608.01349, *1608.02362, *Crivellin:2016ejn, *1701.05825, *1703.06019, *1704.06005}. In this letter, we propose a complete model which gives rise to a type of $U(1)$ symmetry that can accommodate the observed low-energy phenomenology.

The characteristics of the new physics that might be responsible for the observed discrepancy with the Standard Model follow quite simply from the particles involved in the decay: a new interaction that {\it (i)} involves both quarks and leptons  and {\it (ii)} has a non-trivial structure in flavour space.
This profile is fit by well-motivated theories that unify quarks and leptons {\it and} have a gauged horizontal~\cite{Maehara:1978ts,*Yanagida:1979gs} --i.e. flavour-- symmetry to address points {\it (i)} and {\it (ii)} respectively.

Let us address first the latter point, that is, horizontal symmetries.
Given the representations of the five SM fermion fields --$q_L\,,u_R\,,d_R\,,\ell_L\,,e_R$-- under the non-abelian part ($SU(3)_c\times SU(2)_L$) of the gauge group, for one family of fermions there is only a single abelian charge assignment possible for a gauge symmetry. 
This is precisely $U(1)_Y$, hence the Standard Model local symmetry, $\mathcal G_{SM}=SU(3)_c\times SU(2)_L\times U(1)_Y$. 
On the other hand, a global $U(1)_{B-L}$ has only a gravitational anomaly; promoting $B-L$ to be gravity-anomaly free {\it and } a local symmetry can be done in one stroke by introducing right-handed (RH) neutrinos, otherwise welcome to account for neutrino masses~\cite{Minkowski:1977sc, *Yanagida:1979as, *Glashow:1979nm, *GellMann:1980vs} and baryogenesis through leptogenesis~\cite{Fukugita:1986hr}. 
The `horizontal' direction of flavour has, on the other hand, three replicas of each field and the largest symmetry in this sector is then $SU(3)^6$. 
Anomaly cancellation without introducing any more fermion fields nevertheless restricts the symmetry which can be made local to $SU(3)_Q\times SU(3)_L$. 
It is worth pausing to underline this result; {\it the largest anomaly-free local symmetry extension that the SM}+3$\nu_R$ {\it admits is} $ SU(3)_Q\times SU(3)_L\times U(1)_{B-L}$. 
However, now turning to point $(i)$, one realises that the horizontal symmetries above do not connect quarks and leptons in flavour space. 
Although it is relatively easy to break the two non-abelian groups to the diagonal to satisfy {\it (i)}, the desired structure can arise automatically from a unified theory; one is then naturally led to a Pati-Salam~\cite{Pati:1974yy} model $SU(4)\times SU(2)_L\times SU(2)_R\times SU(3)_H$, which also solves the Landau pole problem of $U(1)_{B-L}$ and $U(1)_Y$. 

Explicitly:
\begin{align}
  \mathcal G&=SU(4)\times SU(2)_L\times SU(2)_R\times SU(3)_H\\
  \psi_L&=\left(\begin{array}{cc} u_L&d_L \\ \nu_L & e_L  \end{array}\right)\,\qquad
  \psi_R=\left(\begin{array}{cc} u_R&d_R\\ e_R&\nu_R  \end{array}\right)\,
  \label{PSHModel}
\end{align}
where $\psi_L\sim(4,2,1)$ and $\psi_R\sim(4,1,2)$ under Pati-Salam, and both are in a fundamental representation of $SU(3)_H$.

The breaking of the Pati-Salam group, however, occurs differently from the usual $SU(4)\times SU(2)^2\to \mathcal G_{SM}$; instead we require $SU(4)\times SU(2)^2\to \mathcal G_{SM}\times U(1)_{B-L}$. 
This can be done breaking separately $SU(4)\to SU(3)_c \times U(1)_{B-L}$ and $SU(2)_{R}\to U(1)_{3}$ with $U(1)_3$ being right-handed isospin --we recall here that hyper-charge is $Q_Y=Q_{B-L}/2+\sigma_3^R$. 
This breaking would require two scalar fields in each sector to trigger the breaking; the detailed discussion of this mechanism nevertheless is beyond the scope of this work and will not impact the low energy effective theory.

\section{The Model}
Having discussed the Pati-Salam motivation for our horizontal symmetry, we shall now walk the steps down to the low energy effective theory and the connection with the SM.
At energies below unification yet far above the SM scale we have the local symmetry:
\begin{align}
  \mathcal G=\mathcal G_{SM}\times SU(3)_H\times U(1)_{B-L}\,.
\end{align}
The breaking $SU(3)_H \times U(1)_{B-L} \to U(1)_h$ occurs as one goes down in scale with the current of the unbroken symmetry being:
\begin{gather}
  J_\mu^{h}=\bar \psi \gamma_\mu \left(  g_H c_\theta T^H_{\mathbb CS}+g_{B-L}s_\theta Q_{B-L} \right)
  \equiv g_h \bar \psi \gamma_\mu T^h_\psi \psi\\
  T^h_\psi=  T^H_{\mathbb CS}+t_\omega Q_{B-L} \,, 
\end{gather}
where $T^{H}_{\mathcal CS}$ is an element of the Cartan sub-algebra of $SU(3)$, i.e. the largest commuting set of generators (which we can take to be the diagonal ones), $\psi$ is the Dirac fermion $\psi_L+\psi_R$ with the chiral fields given in Eq.~(\ref{PSHModel}), and $\theta$ is an angle given by the representation(s) used to break the symmetry. 
Before proceeding any further, it is useful to give explicitly the basis-invariant relations that the generators of this $U(1)_h$ satisfy:
\begin{align}
  \mbox{Tr}_{\text{fl}}&\left( T^hT^h\right)=\frac{1}{2}+3t_\omega^2 Q_{B-L}^2\,, \\
  \mbox{Tr}_{\text{fl}}&\left( T^h\right)=3t_\omega Q_{B-L}\,,
\end{align}
where the trace is only over flavour indices, there is a generator $T^h$ for each fermion species including RH neutrinos, and the sign of the traceless piece of $T^h$ is the same for all fermion representations.

The one condition we impose on the flavour breaking $SU(3)_H\times U(1)_{B-L} \to U(1)_h$ is that the unbroken $U(1)_h$ allows for a Majorana mass term for RH neutrinos, such that they are heavy and can give rise to leptogenesis and small active neutrino masses via the seesaw formula. 
A high breaking scale is further motivated by the need to suppress FCNC mediated by the $SU(3)_H$ gauge bosons. 
The desired breaking pattern can be achieved by introducing fundamental $SU(3)_H$ scalar fields, which at the same time generate the Majorana mass term. 
Let us briefly sketch this: we introduce two scalars\footnote{These scalars can each be embedded in a $(4,1,2)$ multiplet under the $SU(4)\times SU(2)_L\times SU(2)_R$ Pati-Salam group.} $\phi_1$, $\phi_2$ in $(3,-1)$ of $SU(3)_H\times U(1)_{B-L}$, so that we can write:
\begin{align}
  \bar\nu^c_R \lambda_{ij} \phi_i^*\phi_j^\dagger \nu_R+h.c.
\end{align}
This implies two generations of RH neutrinos have a large Majorana mass ($\sim 10^{10}$GeV), which is the minimum required for leptogenesis~\cite{hep-ph/0208157} and to produce two mass differences for the light neutrinos $\nu_L$  --one active neutrino could be massless as allowed by data.
The third RH neutrino requires an extra scalar field charged under $U(1)_h$ to get a mass; depending on the charge of the scalar field this might be a non-renormalisable term, making the RH neutrino light and potentially a dark matter candidate.

The second role of these scalar fields is symmetry breaking; in this sense two fundamentals of an $U(3)$ symmetry can {\it at most} break it to $U(1)$, this makes our $U(1)_h$ come out by default.
To be more explicit, with all generality one has $\left\langle \phi_1 \right\rangle =(v_H,0,0), \left\langle \phi_2 \right\rangle =v'_H(c_\alpha,s_\alpha,0)$ and then for $s_\alpha\neq 0$ there is just one unbroken $U(1)$ whose gauge boson $Z_h$ is the linear combination that satisfies:
\begin{align}\label{U1hcond}
  D_\mu \left \langle \phi_{1,2}\right\rangle=\left(g_H T A^H_\mu- g_{B-L} A^{B-L}_\mu \right)\left \langle \phi_{1,2}\right\rangle=0 \,.
\end{align}
Given the v.e.v. alignment, the solution involves  $T_8$  in $SU(3)_H$ and via the rotation $A^{H,8}=c_\theta Z_h -s_\theta A^\prime, A^{B-L}=s_\theta Z_h+c_\theta A^\prime$, where $A^\prime$ is the massive gauge boson, we find that the solution to Eq.~(\ref{U1hcond}) is:
\begin{align}
  t_\theta=&\frac{1}{2\sqrt3}\frac{g_H}{ g_{B-L}}\,,& t_\omega&=t_\theta\frac{g_{B-L}}{g_H}=\frac{1}{2\sqrt3}\,,
\end{align}
with $g_h=g_Hc_\theta$, in close analogy with SM electroweak symmetry breaking (EWSB). 
This solution implies, for leptons
\begin{align}\label{ThLeptons}
  T^h_L =& T_{8}^H-t_\omega \mathbb 1 =  \frac{ 1}{2\sqrt{3}} \left( \begin{array}{ccc}0&&\\&0&\\&&-3 \end{array} \right), 
\end{align}
whereas for quarks
\begin{align} \label{ThQuarks}
  T^h_Q=& T^H_{8}+\frac{1}{3}t_{\omega}\mathbb 1=\frac{1}{2\sqrt3}\left(\begin{array}{ccc}\frac43&&\\&\frac43&\\&&-\frac53\end{array}\right)\,.
\end{align}

At this level the current that the $U(1)_h$ couples to is different for quarks ($T^h_Q$) and leptons ($T^h_L$) but \emph{vectorial} for each of them. 
On the other hand, most previous $Z'$ explanations for the LFU anomalies have considered phenomenologically motivated \emph{chiral} $U(1)$ symmetries. Of course, the above charge assignment is one of several possibilities that can be obtained from a bottom-up approach\footnote{Additional assumptions on the rotation matrices in~\cite{BCD} lead to different mass-basis couplings from those we consider.}~\cite{BCD}; however, as we have shown, this particular flavour structure is well-motivated by the underlying UV theory.

The last step to specify the low energy theory is to rotate to the mass basis of all fermions. 
In this regard some comments are in order about the explicit generation of masses and mixings in this model.
Charged fermion masses would require the introduction of scalar fields charged under both the electroweak and the horizontal group.\footnote{Alternatively, the effective Yukawa couplings can be generated by assuming a horizontal singlet Higgs doublet at the electroweak scale and introducing two pairs of Dirac fermions for each of the six fermion fields, $q_L$, $u_R$, $d_R$, $l_L$, $e_R$ and $\nu_R$ at the $SU(3)_H$ breaking scale, and one pair of these fermions at the $U(1)_h$ breaking scale. The extra fermions are all $SU(3)_H$ singlets. See \cite{1106.1734} for a similar mechanism.} 
At scales above the $U(1)_h$ breaking the fields can be categorised according to their $U(1)_h$ charge\footnote{Ultimately, these three Higgs belong to $H(2,1/2,{\bf8})$ and $H(2,1/2,{\bf1})$ under $SU2_L\times U(1)_Y\times SU(3)_H$. To realise mass matrices for the quarks and leptons requires three $H(2,1/2,{\bf8})$ and one $H(2,1/2,{\bf1})$ at the scale of $G_{SM}\times SU(3)_H\times U(1)_{B-L}$.}; one would need at least a charge $3$, a charge $-3$ and a neutral --in units of $g_h/2\sqrt3$-- `Higgs' transforming as $(2,1/2)$ under $SU(2)\times U(1)_Y$; a linear combination of these three much lighter than the rest would emerge as the SM Higgs doublet.
 
An additional SM singlet scalar is also required to break $U(1)_h$ and should simultaneously generate a Majorana mass for the third RH neutrino. If this scalar has $U(1)_h$ charge 3, such a term is non-renormalisable and if suppressed by a unification-like scale yields a keV mass, which is interestingly in a range where this fermion could be dark matter~\cite{1006.1731}. Alternatively, a charge 6 scalar would generate a mass of order a few TeV.

The main focus of this work is, however, the effect of the gauge boson associated with the $U(1)_h$.
In this sense, however generated, the change to the mass basis implies a {\it chiral} unitary rotation. This will change the vectorial nature of the current to give a priori eight different generators $T^h_f$ for each of the eight chiral fermion species after EWSB: $f=u_R\,,u_L\,,d_L\,,d_R\,,\nu_R\,,\nu_L\,, e_L\,,e_R$.
However, before performing the chiral rotations, it is good to recall that the vectorial character of the interaction is encoded in the basis-invariant relations:
\begin{align}\nonumber
  \mbox{Tr}_{\text{fl}}&\left( T^h_fT^h_f\right)=\frac{1}{2}+\frac{1}{4} Q_{B-L}^2\,, \\
  \mbox{Tr}_{\text{fl}}&\left( T^h_f\right)=\frac{\sqrt{3}}{2} Q_{B-L}\,, \label{TrsU1h}
\end{align}
which applies to both chiralities of each fermion field $f$.

As mentioned before a priori all fields rotate when going to the mass basis $f=U_{f}f'$, however we only have input on the mixing matrices that appear in the charged currents: $V_{CKM}=U^\dagger_{u_L} U_{d_L}$ and $U_{PMNS}=U_{e_L}^\dagger U_{\nu_L}$, which involve only LH fields. 
Hence, for simplicity, we assume that RH fields are in their mass bases and need not be rotated.
The CKM matrix is close to the identity, whereas the lepton sector possesses nearly maximal angles; following this lead we assume the angles in $U_{u_L}\,,U_{d_L}$ are small so that there are no large cancellations in $U_{u_L}^\dagger U_{d_L}$, whereas $U_{e_L}$ and $U_{\nu_L}$ have large angles. 
Phenomenologically however, not all angles can be large in $U_{e_L}$ since they would induce potentially fatal $\mu-e$ flavour transitions. 
Hence we restrict $U_{e_L}$ to rotate only in the $2-3$ sector, which could therefore contribute the corresponding factor in the PMNS as suggested in~\cite{hep-ph/0206300}. 
In the quark sector we assume for simplicity that all mixing arises from $U_{d_L}$.
To make our assumptions explicit:
\begin{align}\nonumber
  U_{e_L}&=R^{23}(-\theta_l), & U_{\nu_L}&=R^{23}(\theta_{23}-\theta_l)R^{13}(\theta_{13})R^{12}(\theta_{12}), \\
  U_{u_L}&=\mathbb{1}, & U_{d_L}&=V_{CKM}, \label{MixChoice}
\end{align}
where $R^{ij}(\theta_{ab})$ is a rotation matrix in the $ij$ sector with angle $\theta_{ab}$. 
Hence,
\begin{align}
  T^{h\prime}_{f_L}&= U^\dagger_{f_L} T_f^h U_{f_L}, &
  T^{h\prime}_{f_R}&=T^h_{f_R}, \label{ThMbasis}
\end{align}
and the current reads:
\begin{align}\label{JhPhen}
  J^h_{\mu}=g_h\sum_f \left(\bar f \gamma_\mu T^{\prime h}_{f_L} f_L +\bar f T^{\prime h}_{f_R}f_R \right).
\end{align}
We have now made all specifications to describe the interactions of $Z_h$; all in all only two free parameters, $\theta_l$ and $g_h$, control the couplings to all fermion species.
For those processes well below the $Z_h$ mass ($\sim$TeV), the effects are given at tree level by integrating the $Z_h$ out:
\begin{align}
  S&=\int d^4x  \left\{ \frac12Z_h^\mu \left(\partial^2+M^2 \right)Z_{h,\mu}-g_h Z_h^\mu J^h_\mu  \right\}\\
  &\stackrel{\text {On-shell} \,Z_h}{=}\int d^4x\left(-\frac{1}{2}\frac{g_h^2J_h^2}{M^2}+\mathcal O\left(\partial^2/M^2\right)\right)
\end{align}
with $J^h_\mu$ as given in~(\ref{ThLeptons},\,\ref{ThQuarks},\,\ref{MixChoice}-\ref{JhPhen}), so that the effective action depends on $\theta_l$ and $M/g_h$.

\section{Low Energy Phenomenology}
The most sensitive probes of $Z_h$ effects come from flavour observables, in particular the FCNC produced in the down sector. 
An important consequence of the rotation matrices in Eq.~\eqref{MixChoice} is that these FCNC have a minimal flavour violation (MFV)~\cite{Chivukula:1998vd,DAmbrosio:2002vsn} structure: $\bar d^i\gamma_\mu V_{ti}^*V_{tj}d_j$.  
Additionally, there can be charged lepton flavour violation (LFV) involving the $\tau-\mu$ transition. 
Even after allowing for these constraints, the $Z_h$ could also potentially be accessible at the LHC. 
Effects on other potentially relevant observables including the muon $g-2$, $Z$-pole measurements at LEP, and neutrino trident production are sufficiently suppressed in our model.
Below we discuss the relevant phenomenology in detail.

\subsection{Semi-leptonic $B$ decays}
The relevant Lagrangian for semi-leptonic $B_s$ decays is
\begin{align}
  \mathcal L_{B_s}=-\frac{3}{4}\frac{g_h^2}{M^2}\left(V_{tb}V_{ts}^*\,\bar s\gamma_\mu  b_L\right)\left(J^\mu_{l_L}+J^\mu_{l_R}+J^\mu_{\nu_L} \right)+h.c.\,,
\end{align}
where for simplicity we have assumed all three RH neutrinos are not accessible in $B$ decays and we have
\begin{align}
J_{l_L}^\rho&=s_{\theta_l}^2\bar \mu\gamma^\rho \mu_L+c_{\theta_l}^2\bar\tau \gamma^\rho \tau_L+s_{\theta_l}c_{\theta_l}\bar \mu\gamma^\rho \tau_L+h.c.\,,\\
J_{l_R}^\rho&=\bar\tau_R\gamma^\rho \tau_R\,,\\ \label{RHLCurr}
J_{\nu_L}^\rho&=\bar\nu^i \gamma^\rho (U_{\nu_L}^*)_{3i} (U_{\nu_L})_{3j}  \nu_L^j\,.
\end{align}

In recent times, a number of measurements of $b\rightarrow s\mu\mu$ processes have shown discrepancies from their SM predictions, most notably in the theoretically clean LFU violating ratios $R_K$ and $R_K^*$.
Global fits to LFU violating data suggest that the observed discrepancies can be explained via a new physics contribution to the Wilson coefficients $C_{9,10}^l$, with the preference over the SM around $4\sigma$~\cite{1704.05435,1704.05438,1704.05340,1704.05444,1704.05447,Geng:2017svp}. 
The effective Hamiltonian is defined as
\begin{equation}
  \mathcal{H}_{\text{eff}} = -\frac{4G_F}{\sqrt{2}}V_{tb}V_{ts}^*\left(C_9^l\mathcal{O}_9^l+C_{10}^l\mathcal{O}_{10}^l+C_\nu\mathcal O_\nu\right) \,,
\end{equation}
where
\begin{align}
  \mathcal{O}_9^l&=\frac{\alpha}{4\pi}\left(\bar s\gamma_\mu \,b_L\right)\left(\bar l\gamma_\mu l \right), \\ 
  \mathcal{O}_{10}^l&=\frac{\alpha}{4\pi}\left(\bar s\gamma_\mu \,b_L\right)\left(\bar l\gamma_\mu\gamma^5 l \right),\\
  \mathcal{O}_\nu^{ij}&= \frac{\alpha}{2\pi}\left(\bar s\gamma_\mu \,b_L\right)\left(\bar \nu^i \gamma_\mu \nu_L^j \right)\,.
\end{align}
In our model, separating the Wilson coefficients into the SM contribution ($C_{SM}$) and the $Z_h$ piece ($\delta C$) we have, for muons:
\begin{equation}
  \delta C_9^\mu=-\delta C_{10}^\mu=-\frac{\pi}{\alpha\sqrt{2}G_F}\frac{3}{4}\frac{g_h^2}{M^2}s_{\theta_l}^2 \,.
\end{equation}
In fitting the observed anomalies we use the results of Ref.~\cite{1704.05435}, which for the relevant scenario $\delta C_9^\mu=-\delta C_{10}^\mu$ gives $\delta C_9^\mu\in[-0.81\,-0.48]$ $([-1.00,\,-0.34])$ at $1(2)\sigma$. 
The fully leptonic decay $B_s\rightarrow\mu\mu$ provides an additional constraint on $\delta C_{10}^\mu$; the current experimental value~\cite{1411.4413} is consistent with the above best-fit region.

There is also a contribution to decays involving neutrinos, $B\rightarrow K^{(*)}\nu\bar\nu$, where we now have:
\begin{align}
\delta C_{\nu}^{ij}&=\delta C_\nu (U_{\nu_L}^*)_{3i} (U_{\nu_L})_{3j}\,,& \delta C_\nu&=-\frac{\pi}{\alpha\sqrt{2}G_F}\frac{3}{4}\frac{g_h^2}{M^2}\,,
\end{align}
so that the ratio to the SM expectation reads:
\begin{equation}
  R_{\nu\bar\nu}\equiv\frac{\Gamma}{\Gamma_\text{SM}}=1+\frac{2}{3}\left(\frac{\delta C_\nu}{C^\nu_{SM}}\right)+\frac{1}{3}\left(\frac{\delta C_\nu}{C^\nu_{SM}}\right)^2 \,,
\end{equation}
where $C^\nu_{SM}\approx-6.35$~\cite{1409.4557}. 
Notice that this is independent of the mixing in the lepton sector, and the rate is always enhanced.
The current experimental bound on this ratio is $R_{\nu\bar\nu}<4.3$ at 90\% CL~\cite{1303.3719,1303.7465}.

Depending on the mixing angle in the lepton sector, the SM-background free LFV decay $B\rightarrow K^{(*)}\tau\mu$ can also be significantly enhanced, 
whereas there is an irreducible contribution to $B\to K^{(*)}\tau\tau$ from the RH currents in Eq.~(\ref{RHLCurr});
both of these contributions nevertheless lie well below the current experimental bounds~\cite{1504.07928,Flood:2010zz}.

Finally, one might also expect similar contributions in $b\rightarrow d$ and $s\rightarrow d$ transitions, the latter leading to effects in $K$ decays. 
However, given our assumptions on the mixing matrices, the MFV structure in the down quark couplings means that these contributions are sufficiently suppressed. 
In particular, the otherwise stringent bound from $K\to\pi\nu\bar{\nu}$~\cite{0808.2459, hep-ph/0603079} is found to be comparable, yet still sub-dominant, to that from $B\to K\nu\bar{\nu}$.

\subsection{$\bar B\text{--}B$ Mixing}

The $Z_h$ gives a tree-level contribution to $\bar B_s\text{--}B_s$ and $\bar B_d\text{--}B_d$ mixing, which provide some of the most stringent constraints on the model. The relevant Lagrangian is 
\begin{align}
  \mathcal L_{\Delta B=2}=-\frac{3}{8}\frac{g_h^2}{M^2}\left(V_{tb}V_{ti}^*\,\bar d_i\gamma_\mu  b_L\right)^2\,.
\end{align}
This leads to a correction to $\Delta m_B$ given by
\begin{equation}
  C_B\equiv\frac{\Delta m_{B}}{\Delta m_B^\text{SM}}=1+\frac{4\pi^2}{G_F^2m_W^2\hat{\eta}_BS(m_t^2/m_W^2)}\frac{3}{8}\frac{g_h^2}{M^2}c(M) \,,
\end{equation}
where $S(m_t^2/m_W^2)\approx2.30$ is the Inami-Lim function~\cite{Inami:1980fz}, $\hat{\eta}_B\simeq0.84$ accounts for NLO QCD corrections~\cite{Buras:1990fn,1008.1593}, and $c(M)\approx0.8$ includes the running from $M$ down to $m_B$ using the NLO anomalous dimension calculated in Refs.~\cite{hep-ph/9711402, hep-ph/0005183}.
This observable is tightly constrained, yielding $0.899<C_{B_s}<1.252$ and $0.81<C_{B_d}<1.28$ at 95\% CL~\cite{0707.0636}. 

Once again, the MFV structure of the couplings ensures that effects in $\bar K\text{--}K$ mixing are well below current bounds. 
In this case the SM prediction for $\Delta m_K$ also suffers from theoretical uncertainties.

\subsection{Lepton Flavour Violation in $\tau\to \mu$}
There is a contribution to the cLFV decay $\tau\to 3 \mu$:
\begin{align}
  \mathcal{L}_{\text{LFV}} = -\frac{3}{4}\frac{g_h^2}{M^2} s_{\theta_l}^3 c_{\theta_l}\, \bar \tau\gamma^\rho \mu_L \,\bar\mu\gamma_\rho \mu_L \,,
\end{align}
resulting in a branching ratio 
\begin{equation}
  \text{BR}(\tau\rightarrow 3\mu)=\frac{m_\tau^5}{1536\pi^3\Gamma_\tau}\frac{g_h^4}{M^4}\frac{9}{8}s_{\theta_l}^6c_{\theta_l}^2 \,.
\end{equation}
The current experimental bound is $\text{BR}(\tau\rightarrow 3\mu)<2.1\times10^{-8}$ at 90\% CL~\cite{1001.3221}. 
This restricts the allowed values of the mixing angle $\theta_l$.

\subsection{Collider Searches}

Depending on its mass, the $Z_h$ may be directly produced at the LHC. 
The large $U(1)_h$ charge in the lepton sector results in a potentially sizeable branching ratio into muons: BR$(Z_h\rightarrow\mu\mu)\simeq0.08\,s_{\theta_l}^4$. 
The strongest bounds on a spin-1 di-muon resonance are from the ATLAS search at $\sqrt{s}=13\,$TeV with 36$\,\text{fb}^{-1}$~\cite{ATLAS-CONF-2017-027}. 
Furthermore even for very large masses, $M\gtrsim6\,$TeV, non-resonant production will continue to provide bounds; these can become important in the future~\cite{1704.09015}. 
Di-jet searches also provide a complementary strategy, although the constraints are weaker.

\subsection{Perturbativity}

The one-loop beta function for $U(1)_h$ is 
\begin{equation}
  \beta(g_h)=\frac{269}{36}\frac{g_h^3}{(4\pi)^2}\,,
\end{equation}
where we have assumed the $U(1)_h$ breaking scalar has charge $3$.
The gauge coupling $g_h$ then encounters a Landau pole at the scale
\begin{equation}
  \Lambda=\exp\left(\frac{288\pi^2}{269g_h(M)^2}\right)M \,.
\end{equation}
This scale should at least be larger than the $SU(3)_H \times U(1)_{B-L} \to U(1)_h$ breaking scale. 
Assuming that the breaking occurs at $10^{10}\,$GeV --so that the RH neutrinos obtain a sufficiently large mass for viable leptogenesis-- leads to the bound $g_h(10\,\text{TeV})\lesssim0.9$. 
Also note that depending on the specific UV mechanism for generating the fermion mass matrices, $SU(3)_H$ may not remain asymptotically free, in which case there can be additional constraints from perturbativity.

\section{Discussion}

\begin{figure}[htbp]
  \centering
  \includegraphics[width=.4\textwidth]{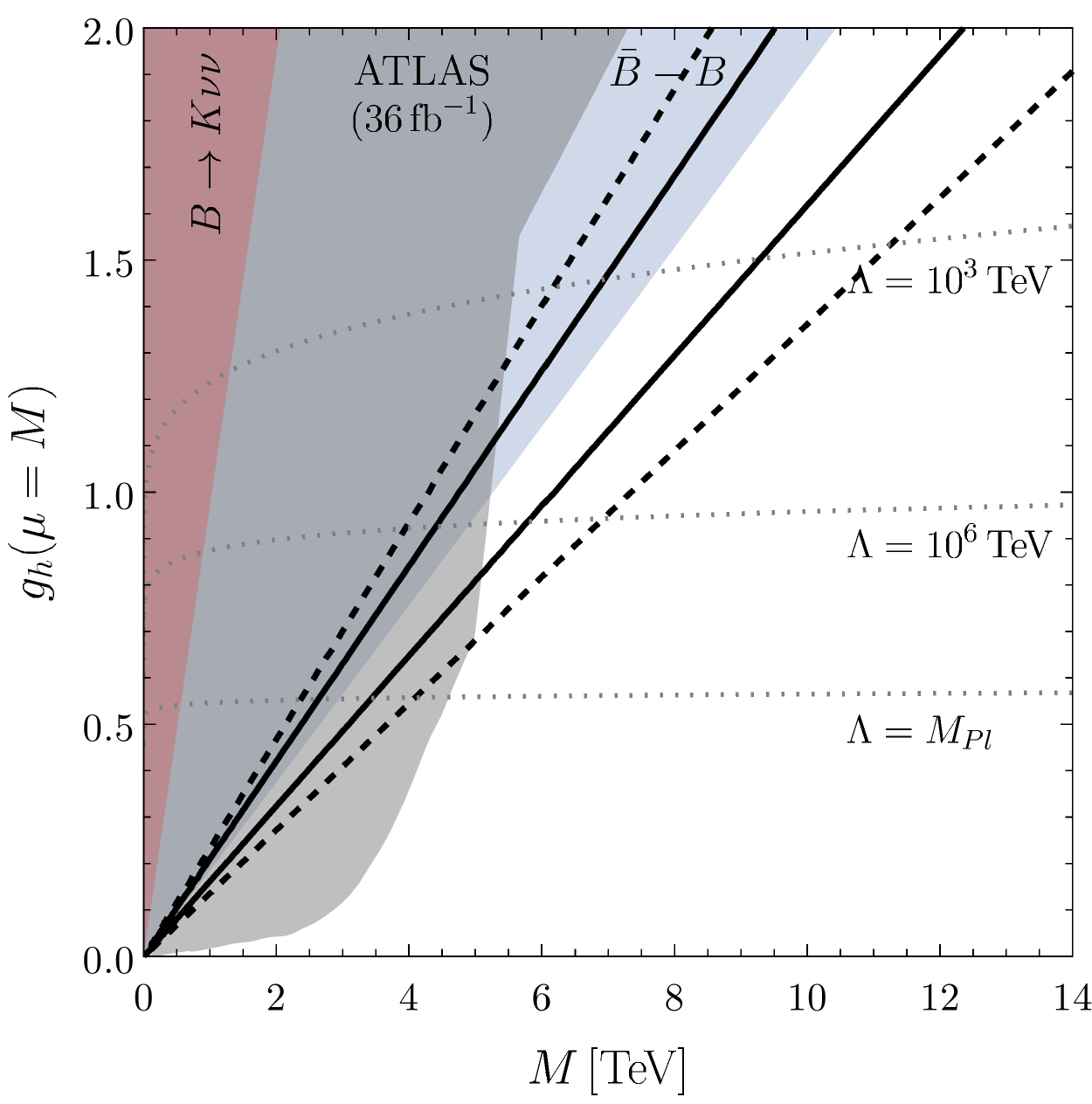}
  \caption{The best-fit region to the LFU anomalies at $1\sigma$ (solid lines) and $2\sigma$ (dashed lines). The shaded regions are excluded by existing measurements at 95\% CL. The dotted lines correspond to upper bounds on the $SU(3)_H\times U(1)_{B-L}$ breaking scale from perturbativity. We have fixed $\theta_l=\pi/2$.}
  \label{fig:M-g_h}
\end{figure}

In Fig.~\ref{fig:M-g_h} we combine the above constraints and show the region of parameter space which can explain the observed LFU anomalies. 
It is clear that this scenario is already tightly constrained by the existing measurements, in particular $\bar B\text{--}B$ mixing and LHC searches. 
Requiring perturbativity up to the scale of the right-handed neutrinos ($\gtrsim10^{10}\,$GeV) provides an additional upper bound on the gauge coupling, leaving a small region of parameter space consistent with the best fit value of $C_9^\mu$ at $1\sigma$. 
The $2\sigma$ region for $C_9^\mu$ --still a significant improvement over the SM-- opens up substantially more viable parameter space. 

\begin{figure}[htbp]
  \centering
  \includegraphics[width=.4\textwidth]{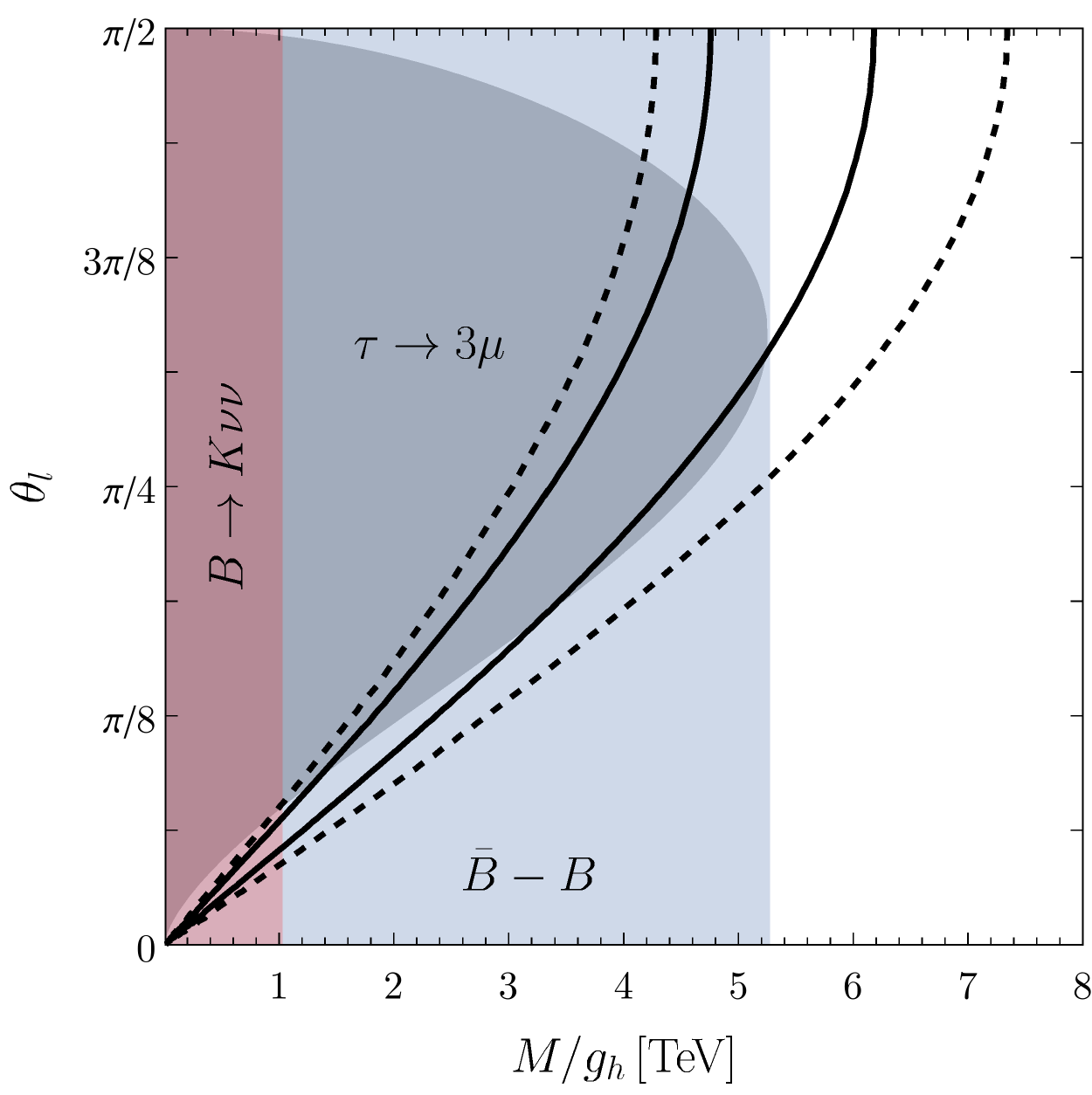}
  \caption{Same as Fig.~\ref{fig:M-g_h}, but showing the dependence on $\theta_l$.}
  \label{fig:M-theta_l}
\end{figure}

The dependence on the mixing angle in the lepton sector is shown in Fig.~\ref{fig:M-theta_l}. 
Consistency with the $2\sigma$ best-fit region for the anomalies and the bounds from $\bar B\text{--}B$ mixing requires $\theta_l\gtrsim\pi/4$. 
There is also a potentially important additional constraint from $\tau\rightarrow3\mu$. 
In the $M-g_h$ plane, the situation remains similar to Fig.~\ref{fig:M-g_h}; however the best-fit regions for the anomaly move towards smaller masses as $\theta_l$ is reduced. 
Let us also comment briefly on the mixing in the quark sector. For simplicity, in Eq.~\eqref{MixChoice} we made the assumption $U_{d_L}=V_{CKM}$. 
Allowing instead for an arbitrary angle, one obtains the upper bound $\theta_{23}\lesssim0.08$; this is qualitatively similar to the case we have considered $(|V_{ts}|\simeq0.04)$. 
For $\theta_{23}$ below this value, $\bar B\text{--}B$ mixing can be alleviated, but the bounds from LHC searches and perturbativity become more severe.

One consequence of the relatively strong experimental constraints is that this model can be readily tested in the relatively near future. 
Improved precision for $\Delta m_B$ would either confirm or rule out this model as a potential explanation for the LFU anomalies. 
On the other hand, improvements in the LHC limit, when combined with the perturbativity bounds, would force one to consider lower $SU(3)_H \times U(1)_{B-L} \to U(1)_h$ breaking scales. 
In addition, the LFV decay $\tau\rightarrow3\mu$ provides an important complementary probe of the mixing angle in the lepton sector. 
Similarly, the decay $B\rightarrow K^{(*)}\tau\mu$ can be significantly enhanced and could be observable in the future.
In this sense it is good to note that the vectorial character of the $U(1)_h$ reveals itself in the sum rules
\begin{align}
  \sum_{l}\delta C_{10}^{ll}&=0\,, &
  \sum_{l}\delta C_{9}^{ll}=2\sum_{i} \delta C_{\nu}^{ii}\,,
\end{align}
\begin{align}
  \sum_{ll'}\left(|\delta C_{9}^{ll'}|^2+|\delta C_{10}^{ll'}|^2\right)=4\sum_{ij} |\delta C_{\nu}^{ij}|^2\,,
\end{align}
which is basically a manifestation of Eq.~\eqref{TrsU1h}.

Finally, we have focused on the specific case of a $G_{SM}\times SU(3)_H\times U(1)_{B-L}$ symmetry, however there exist other related scenarios which provide equally interesting possibilities. For example, if one instead assumes $G_{SM}\times SU(3)_Q\times SU(3)_L\times U(1)_{B-L}$, it is possible to obtain $T^h_L\sim$ diag(0,0,-3) and $T^h_Q\sim$ diag(0,0,1). 
This is nothing other than a $U(1)_{B-L}$ under which only the third generation is charged.
The LHC bounds would be significantly weakened in such a scenario; $g_h$ could then remain perturbative up to the Planck scale. 
Another possible symmetry is $G_{SM}\times SU(3)_Q\times SU(3)_L$ if a bifundamental Higgs $(3,\,3^*)$ condenses at low energies, since it mixes two U(1) gauge bosons.
A merit of this model is that one can give heavy Majorana masses to all right-handed neutrinos by taking the unbroken $U(1)_h$ as diag(0,1,-1) for leptons~\cite{1008.4445}, and diag(1,1,-2) for quarks. 
The low energy phenomenology of a $U(1)$ with similar flavour structure was previously considered in~\cite{CDJ,CFGI}, the latter based on another non-abelian flavour symmetry~\cite{Grinstein:2010ve,*Alonso:2016onw}.
We leave the detailed investigation of such related scenarios for future work, but application of our analysis is straightforward.

\section{Conclusion}

If confirmed, the violation of lepton flavour universality would constitute clear evidence for new physics. 
In this letter, we have proposed a complete, self-consistent model in which the observed anomalies are explained by the presence of a new $U(1)_h$ gauge symmetry linking quarks and leptons. 
We have shown how such a symmetry can naturally arise from the breaking of an $SU(3)_H\times U(1)_{B-L}$ horizontal symmetry. 
Furthermore, within the SM+3$\nu_R$, this is the largest anomaly-free symmetry extension that is consistent with Pati-Salam unification. 
The model is readily testable in the near future through direct searches at the LHC, improved measurements of $\bar B\text{--}B$ mixing and charged LFV decays.

\noindent {\bf{Acknowledgements}}
R.A. thanks IPMU for hospitality during the completion of this work.
This work is supported by Grants-in-Aid for Scientific Research from the Ministry of Education, Culture, Sports, Science, and Technology (MEXT), Japan, No. 26104009 (T.T.Y.), No. 16H02176 (T.T.Y.) and No. 17H02878 (T.T.Y.), and by the World Premier International Research Center Initiative (WPI), MEXT, Japan (P.C., C.H. and T.T.Y.). 
This project has received funding from the European Union's Horizon 2020 research and innovation programme under the Marie Sk\l{}odowska-Curie grant agreement No 690575.

\bibliography{SU3HB-L}

\end{document}